# Optimization of Fuzzy Semantic Networks Based on Galois Lattice and Bayesian Formalism

**Mohamed Nazih Omri**

ERPAH-FST & DMI-IPEIM,
Route de Kairouan, 5019 Monastir.
Tel. 216 3 500 273, Fax. 216 3 500 512
E-mail: Nazih.Omri@ipeim.rnu.tn

**Abstract**

This paper presents a method of optimization, based on both Bayesian Analysis technical and Galois Lattice of Fuzzy Semantic Network. The technical System we use learns by interpreting an unknown word using the links created between this new word and known words. The main link is provided by the context of the query. When novice's query is confused with an unknown verb (goal) applied to a known noun denoting either an object in the ideal user's Network or an object in the user's Network, the system infer that this new verb corresponds to one of the known goal. With the learning of new words in natural language as the interpretation, which was produced in agreement with the user, the system improves its representation scheme at each experiment with a new user and, in addition, takes advantage of previous discussions with users. The semantic Net of user objects thus obtained by learning is not always optimal because some relationships between couple of user objects can be generalized and others suppressed according to values of forces that characterize them. Indeed, to simplify the obtained Net, we propose to proceed to an Inductive Bayesian Analysis, on the Net obtained from Galois lattice. The objective of this analysis can be seen as an operation of filtering of the obtained descriptive graph.

## 1 Introduction

In order to respond to a query, an executive assistant might know very precisely the goal the user has in mind, which means an object in a given state (the properties of the object being transformed). Moreover, even when

1844

goals are fairly well defined, it is often necessary to think about superordinate goals. Let's take (example 1) the query of a subject using a Macintosh Computer.

The Galois lattice [6] and the fuzzy set methods have been used to develop the "on-line instructions" mechanisms of an Intelligent Assistance System. It can be seen as a supervisor of task execution that has the "ideal user's knowledge" of (i) prerequisites of procedures, (ii) subGoals structure. And (iii) the semantic network of the elements of the device where applied procedures are used as properties, as well as (iv) the knowledge of perceptible and imperceptible effects of user's actions. With an interactive dialogue with a user, the Assistance System tries to match items provided by users in natural language with the knowledge included in the ideal user's semantic network [7], [12].

The example of the technical system we consider here is Word Processor software (figure1), with Objects such as "chain-of-characters", and procedures such as "cut" or "copy". For a novice user of the software, the list of standard denominations is not obvious and he often would like to ask an expert operator about how to execute an action such as "how to rub letters" [12], [13].

## 2  The Ideal Expert's and Novice User's Fuzzy Semantic Net

We define the ideal Expert knowledge of a system as the knowledge that is sufficient to the system and that is described in a semantic network (figure1). Construction of the Ideal Expert Knowledge starts if given a set of tasks that are executed using elements of one technical device through procedures. The first step is the task decomposition as a hierarchy of Goal decomposition into subGoals from the level of the Goal of the task to primitive actions. The second step consists in (i) drawing up a list of possible Goals and the procedures to reach these Goals (ii) constructing the Ideal Expert Net as a classical semantic network. But, instead of using structural properties of systems interface Objects; Goals reachable with those Objects are used as properties. The ideal user's description uses valid procedures that have to be applied to the elements of the device in order to successfully complete the task. Classes of Objects and relations between classes of Objects merge from routines for classification and routines for classes organization [22].

However given the polysemic aspects of natural language (verbs and nouns which express goals and device objects), with the necessity of a man-machine interface that involve queries of users, the problem that is under investigation is how to match the content of a query (the label of an Object and the label of a Goal applied to this Object, as expressed by a novice user) to their corresponding items (class of Objects and Goals as properties) in the Ideal Expert Net. By answering queries of the users while they try to perform a given goal, the Expert Assistant delivers not only planning information, but also a goal structure and the knowledge of what justifies the procedure by providing the knowledge that is included in the Ideal Expert Net [12]. If the Assistance System does not understand the meaning of an instruction, it discusses with the user until it is able to interpret the query in its own language [14]. With the learning of new words in natural language as the interpretation produced in agreement with the user, the system improves its representation scheme at each experiment with a new user. And, in addition, takes advantage of previous discussions with users. In a first time the standard Objects and recognized by the software are described in a semantic network where goals stand for properties of Objects. And in a second time, as the queries of an user are expressed in natural language and as they correspond more or less to these standard denominations, the system establishes fuzzy connections between its primary knowledge and the new labels of Objects or procedures expressed by the user [12], [16].

The obtained semantic Net of user objects is not always optimal because some relationships between couple of user objects can be generalized and others suppressed according to values of forces that characterize them. Indeed, to simplify the obtained Net, we propose to proceed to an inductive Bayesian analysis on the obtained Net from Galois lattice [8],[25]. The objective of this analysis can be seen as an operation of filtering of the obtained descriptive graph.

## 3 Optimization of the Fuzzy Semantic Nets by Bayesian Analysis

The approach that we present in this paper is established from Procope's formalism [18], [19], based on the Galois lattice method [9] and the Bayesian formalism [1], [2], [5]. The underlying idea is to end to a hierarchical structure of object users allowing having a process of categorization by discrimination and generalization. To end to a hierarchical structure of user objects in the form of a symbolic data table, the method of the Galois lattice is the means that we have adopted to construct the semantic user object system. This construction consists; from a symbolic table of linguistic data (table 2), to construct, in a first time the binary table (crossed system's objets with user objects are obtained by 0 and 1) (table 3), and in a second time, the different implications between each couple of user objects.

To illustrate this method, we propose to construct the semantic user objects Net corresponding to the following symbolic table (table 2). This last allows us to construct the user objects Net with all possible implications between each couple of objects according to the next rule. Let A and B two Objects defined by a set of property $a_i$ with i ∈ [1, n], we have A implies B if and only if ∀ $a_i$ verifying A then $a_i$ verifying also B. To construct this graph, we have used the software GLG (Galois lattice's Graph) developed in Mathematics and Physics Department of the Preparatory Institute to Studies of Engineer of Monastir and that is going to be published later.

The obtained semantic Net of user objects is not always optimal because some relationships between couple of user objects can be generalized and others suppressed according to values of forces that characterize them. Indeed, to simplify the obtained Net, we propose to proceed to an inductive Bayesian analysis on the obtained Net from Galois lattice [8], [25].

The principal objective of this analysis is to find all the possible oriented dependence existing between different user objects: the knowledge of some will determine - it such or such others. To reply to this objective, we have considered the following user objects: *The number, The Sign, The letters, The numbers, The Characters* and *Substantive*. These user objects represent synonymies by novice users to designate the following system's objects: *Char, Word* and *Key* shown in table 2. To determine the different binary relationships between each couple, the analysis consists to study the implicative structure to each couple, then to all implicative structures corresponding to the form of implicative graph (figure 2).

### 3.1 Descriptive Inductive Analysis

From observations realized on each couple of user objects, we have built the following table (table 4) that presents sorting crossed in effective for each pair of user objects. Each places in table 4 represents 768 users of the software that we have put in place. For instance, in the first places, corresponding to the couple of objects *'the Sign'* and *'the number'*, 100 users have used the word *'the Sign'* to each time that they have used the word *'the number'* to designate a system's object. 30 other users have used the word *'the number'* without used the word *'the Sign'*. 85 have used the word *'the Sign'* without using the word *'the number'* and 553 remainder of the total effective have not used neither the word *'the Sign'* nor the word *'the number'* to designate system's object.

For each of these crossed sorting, we calculate the Loevinger's indication H [3], [4]

associated to the four possible error squares. Positive indices are represented in fat (table 4). If we consider the two values-mark $h_{tend}=0,40$ and $h_{quasi}=0,60$ we have to respect next conclusions:

$H < h_{tend}$ absence of q-implication

$h_{tend} \leq H \leq h_{quasi}$ tendency to the q-implication

$H \geq h_{quasi}$ q–implication

The suitable figure 3 shows two possible cases. The first case, constituted following user objects: *Substantive, The number, The Sign, The letters*, and *The numbers*. Positive connection following q-implication from *The number* to *The* Sign. From *The Sign* to *The letters* with tendency to the equivalence, q-implication with equivalence between *The Sign* and *The numbers*, tendency to the q - implication from *The Sign* to *The numbers* and between *The number* and *Substantive* with tendency to the equivalence. The second case constituted by *the number* user objects, *Substantive, The Characters* and *The Sign* presents relationships of q-exclusion and tendency to the q-exclusion.

## 3.2 Processing by Inductive Bayesian Analysis

This stage consists in determine with the help of the IBA [2], [3], observed oriented relationships descriptively that can be certified inductively, among all relationships in order that the indication $H \geq 0,20$. The objective of this analysis can be seen as an operation of filtering of the obtained descriptive graph (figure 3).

In order that, we are going to calculate, to each places in the table 5 above ($H < 0,20$), the inferior credibility limit, for a guarantee - mark $\delta=90$, for the corresponding indication dress $\eta$. To realize these calculations, we have used a recent version of the software IBA-2 developed in the Cognitive Psychology Laboratory of the Paris8 University and that is going to be published later. Results of these calculations are presented in the following table 6. Negative values are not taken in account and therefore it does not appear in table 5. The results of this filtering allow determining relationships that can be generalized, among the totality of observed relationships descriptively.

According to the graph of the figure 4, we can certify on the one hand, a q-implication with tendency to the equivalence between *The Sign* and *the letters* user objects and a q-implication from *The Sign* to *The number*. We can also certify, on the other hand, a tendency to the q-implication from *Substantive* to *The number*. For the implication from *the letters* to *The Characters* and from this last to *the numbers*, we notices that there is an absence of q-implication with tendency to the exclusion.

## 4 Conclusion

Although the approach presented in this paper, that consists of a learning of new word in natural language in a fuzzy semantic Networks, represent a particular methodology to diagnosis the goal query's novice users and allows identifying the unknown novice user request of the share of the device used. This can serve as basis for our research so as to elaborate a general methodology to diagnosis the purpose goal of the subject, applicable to a large diversity of devices. The objective being to find the totality of compatible purposes with actions of the users, the trip of such graphs facilitates grandly the research. The development of this method would have to allow a best approximation of the category of the purpose aimed by the user and best approaches the diagnosis. We think that it would be interesting to strengthen this tool of softening with the notion of similarity between two Objects (respectively two Goals) so as to establish connection between user Object (or Goal) and system Object (or system Goal) in the semantic Net. This makes only increase performances of the system in the course of the identification of user requests.

hierarchies of concepts. *In O. Rival (ed), Ordered Sets*. Boston: Reidel, pp:445-470.

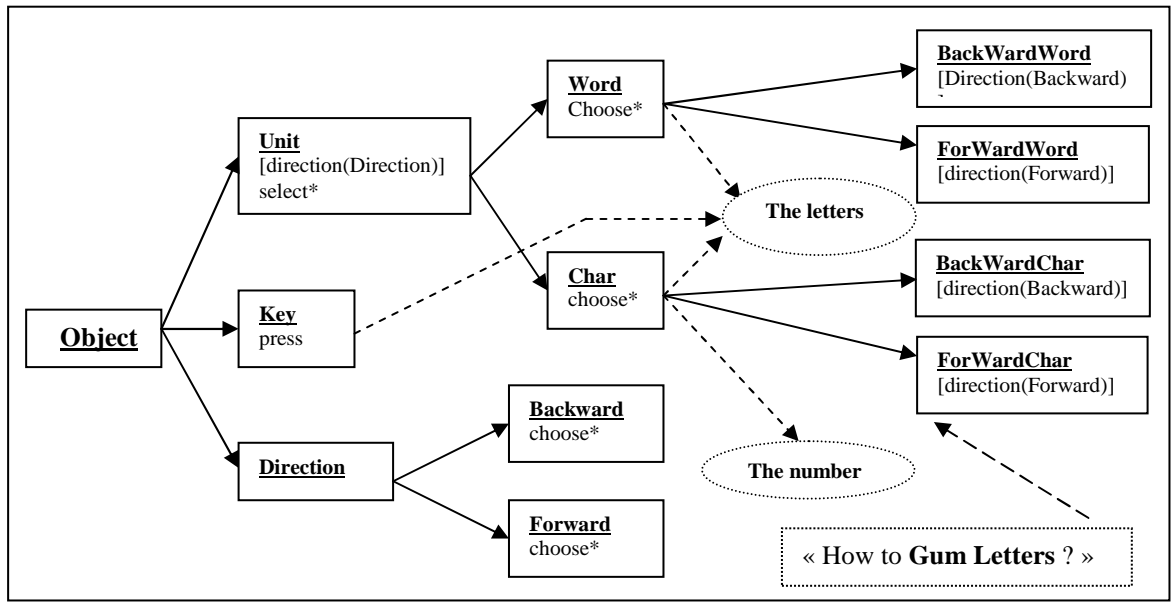

**Figure 1:** The Semantic Network of Novice Users.

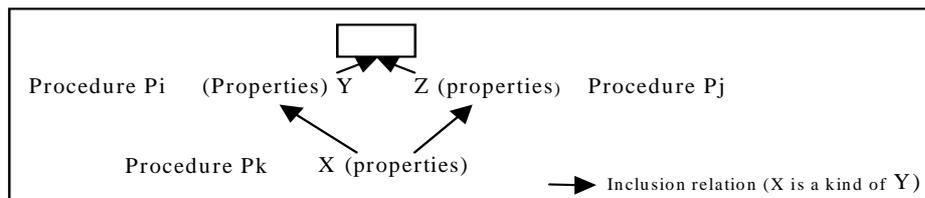

**Figure 2:** *Procedural Semantic Net representation with inclusion relations. Procedural and declarative semantics of the device merges in regard of applied procedures. Classes **Y** and **Z** inherit of procedures of superordinate classes as class **X** inherits of procedures of both **Y** and **Z** classes (multiple inheritance).*

**Table 1:**

|  | Key | Forward-Word | Backward-Word | Forward-Char | Backward-Char | Char | Word | Unit | Direction |
|---|---|---|---|---|---|---|---|---|---|
| Direction (Forward) |  | X |  | X |  | X | X | X |  |
| Direction (Backward) |  |  | X |  | X | X | X | X |  |
| Choose |  | X | X | X | X | X | X |  | X |
| Select |  | X | X | X | X | X | X | X |  |
| Press | X |  |  |  |  |  |  |  |  |

**Table 2.** Example of symbolic table.

|  | Novice User 1 | Novice User 2 | Novice User 3 | Novice User 4 | Novice User 5 |
|---|---|---|---|---|---|
| Char | The number | The Sign | The letters | The numbers | The number |
| Word | The numbers | The letters | Substantive | The Sign | The Sign |
| Key | The Characters | Substantive | Substantive | The Characters | The letters |

**Table 3.** Galois lattice corresponding to the table 2.

|  | The number | The Sign | The letters | The numbers | The Characters | Substantive |
|---|---|---|---|---|---|---|
| Char | 1 | 1 | 1 | 1 | 0 | 0 |
| Word | 0 | 1 | 1 | 1 | 0 | 1 |
| Key | 0 | 0 | 1 | 0 | 1 | 1 |

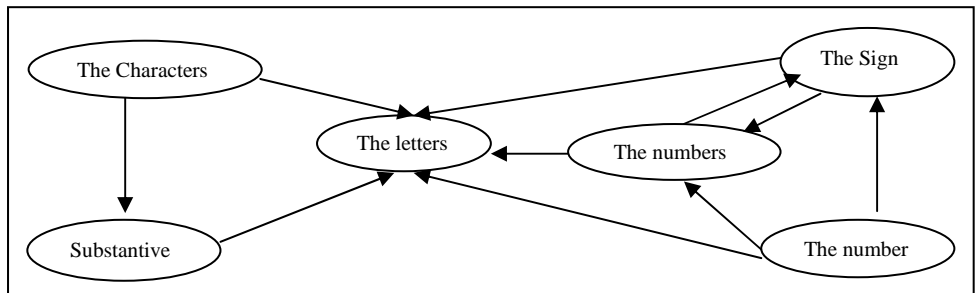

**Figure 3**. The user's objets Net corresponding on the table 2.

**Table 4:** Table of staffs crossed to each couple of user objects.

|            | The Sign |     | The letters |     | The numbers |     | The Characters |     | Substantive |     |
|------------|----------|-----|-------------|-----|-------------|-----|----------------|-----|-------------|-----|
| The number | 100 30   |     | 50 80       |     | 49 81       |     | 38 92          |     | 66 64       |     |
|            | 85  553  |     | 143 495     |     | 100 538     |     | 70 568         |     | 50 588      |     |
| The Sign   |          |     | 150 35      |     | 49 136      |     | 43 142         |     | 46 139      |     |
|            |          |     | 43  540     |     | 100 483     |     | 65 518         |     | 70 513      |     |
| The letters |         |     |             |     | 49 144      |     | 78 115         |     | 26 167      |     |
|            |          |     |             |     | 100 475     |     | 30 545         |     | 90 485      |     |
| The numbers |         |     |             |     |             |     | 49 100         |     | 29 120      |     |
|            |          |     |             |     |             |     | 59 560         |     | 87 532      |     |
| The Characters |      |     |             |     |             |     |                |     | 38 70       |     |
|            |          |     |             |     |             |     |                |     | 78 582      |     |

**Table 5:** Table of Loevinger's indices to each couple of user objects.

|            | The Sign |      | The letters |      | The numbers |      | The Characters |      | Substantive |      |
|------------|----------|------|-------------|------|-------------|------|----------------|------|-------------|------|
| The number | -2,19    | **0,7** | -0,53    | **0,18** | -0,94   | **0,22** | -1,08      | **0,18** | -2,36   | **0,42** |
|            | **0,45** | -0,14 | **0,11** | -0,04 | **0,19**   | -0,05 | **0,22**   | -0,04 | **0,48**   | -0,09 |
| The Sign   |          |      | -2,23      | **0,75** | **0,57** | **0,09** | -0,65   | **0,11** | -0,65    | **0,11** |
|            |          |      | **0,71**   | -0,24 | **0,12**   | -0,03 | **0,21**   | -0,03 | **0,2**    | -0,04 |
| The letters |         |      |            |      | -0,3       | **0,07** | -1,87   | **0,31** | 0,11     | **-0,02** |
|            |          |      |            |      | **0,1**    | -0,02 | **0,18**   | -0,1  | **-0,04**  | 0,01 |
| The numbers |         |      |            |      |             |      | -1,34      | **0,22** | -0,23   | **0,05** |
|            |          |      |            |      |             |      | **0,32**   | -0,05 | **0,07**   | -0,01 |
| The Characters |      |      |            |      |             |      |                |      | -1,33      | **0,24** |
|            |          |      |            |      |             |      |                |      | **0,22**   | -0,04 |

**Table 6 :** Table of inferior credibility limit for each indication H with the guarantee 0.90.

|            | The Sign |      | The letters |      | The numbers |      | The Characters |      | Substantive |      |
|------------|----------|------|-------------|------|-------------|------|----------------|------|-------------|------|
| The number |          | **0,634** |        |      |         | 0,168 |            |      |             | **0,36** |
|            | **0,397** |     |            |      |             |      | 0,156          |      | **0,414**   |      |
| The Sign   |          |      |            | **0,698** |      |      |                |      |             |      |
|            |          |      | **0,658**  |      |             |      | 0,135          |      |             |      |
| The letters |         |      |            |      |             |      |                | **0,264** |         |      |
|            |          |      |            |      |             |      |                |      |             |      |
| The numbers |         |      |            |      |             |      |                | 0,171 |             |      |
|            |          |      |            |      |             |      | **0,253**      |      |             |      |
| The Characters |      |      |            |      |             |      |                |      |             | 0,174 |
|            |          |      |            |      |             |      |                |      | 0,159       |      |

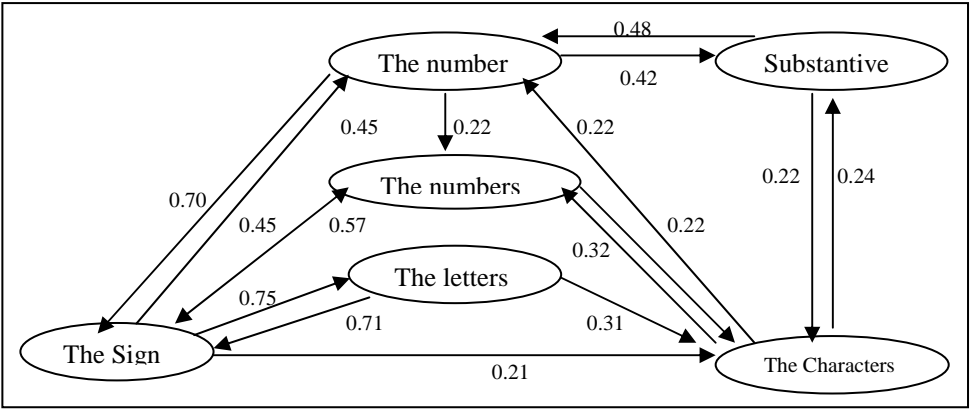

**Figure 4:** The implicative descriptive graph of relationships with the indication H ≥ 0,20.

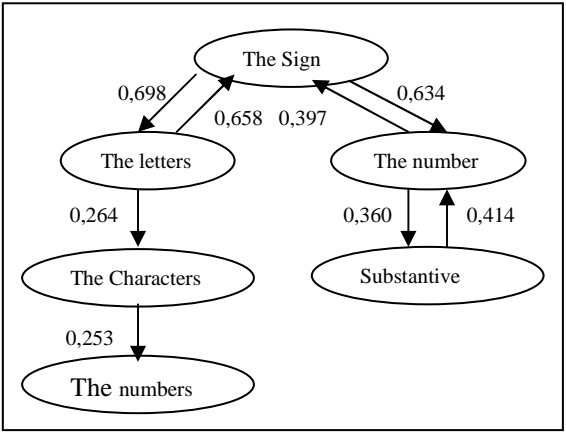

**Figure 5:** The implicative inductive graph of relationships with the indication H ≥ 0,20.